\documentclass[twocolumn,amssymb, amsmath, showpacs, superscriptaddress, aps, prl]{revtex4}
\usepackage{latexsym}        
\usepackage{amssymb}
\usepackage{amsmath}
\usepackage{amsfonts}
\usepackage{graphicx} 
\usepackage{bm}
\usepackage{color}

\begin{document}
\title{Direct observation of nanoscale interface phase in the superconducting chalcogenide K$_{x}$Fe$_{2-y}$Se$_2$ with intrinsic phase separation
}

\author{A. Ricci}
\affiliation{Deutsches Elektronen-Synchrotron DESY, Notkestra$\beta$e 85, D-22607 Hamburg, Germany}
\affiliation{Rome International Center for Materials Science Superstripes RICMASS, via dei Sabelli 119A, 00185 Roma, Italy}

\author{N. Poccia}
\affiliation{MESA+ Institute for Nanotechnology, University of Twente, 7500 AE Enschede, The Netherlands}
\affiliation{Rome International Center for Materials Science Superstripes RICMASS, via dei Sabelli 119A, 00185 Roma, Italy}

\author{B. Joseph}
\affiliation{Dipartimento di Fisica, Universit{\'a} di Roma ``La Sapienza" - P. le Aldo Moro 2, 00185 Roma, Italy}
\affiliation{Elettra-Sincrotrone Trieste, Strada Statale 14, Km 163.5, Basovizza 34149, Trieste, Italy}

\author{D. Innocenti}
\affiliation{Laboratory of Physics of Complex Matter, EPFL - Ecole Polytechnique F{\'e}d{\'e}rale de Lausanne, CH-1015 Lausanne, Switzerland}

\author{G. Campi}
\affiliation{Institute of Crystallography, CNR, via Salaria Km 29.300, Monterotondo Roma, I-00015, Italy}

\author{A. Zozulya}
\affiliation{Deutsches Elektronen-Synchrotron DESY, Notkestra$\beta$e 85, D-22607 Hamburg, Germany}

\author{F. Westermeier}
\affiliation{Deutsches Elektronen-Synchrotron DESY, Notkestra$\beta$e 85, D-22607 Hamburg, Germany}

\author{A. Schavkan}
\affiliation{Deutsches Elektronen-Synchrotron DESY, Notkestra$\beta$e 85, D-22607 Hamburg, Germany}

\author{F. Coneri}
\affiliation{MESA+ Institute for Nanotechnology, University of Twente, 7500 AE Enschede, The Netherlands}

\author{A. Bianconi}
\affiliation{Rome International Center for Materials Science Superstripes RICMASS, via dei Sabelli 119A, 00185 Roma, Italy}

\author{H. Takeya}
\affiliation{National Institute for Materials Science, 1-2-1 Sengen, Tsukuba 305-0047, Japan}

\author{Y. Mizuguchi}
\affiliation{Dipartimento di Fisica, Universit{\'a} di Roma ``La Sapienza" - P. le Aldo Moro 2, 00185 Roma, Italy}
\affiliation{Department of Electrical and Electronic Engineering, Tokyo Metropolitan University, 1-1 Minami-osawa, Hachioji, Tokyo 192-0397, Japan}
\affiliation{National Institute for Materials Science, 1-2-1 Sengen, Tsukuba 305-0047, Japan}

\author{Y. Takano}
\affiliation{National Institute for Materials Science, 1-2-1 Sengen, Tsukuba 305-0047, Japan}

\author{T. Mizokawa}
\affiliation{Dipartimento di Fisica, Universit{\'a} di Roma ``La Sapienza" - P. le Aldo Moro 2, 00185 Roma, Italy}
\affiliation{Department of Complexity Science and Engineering, University of Tokyo, 5-1-5 Kashiwanoha,Kashiwa, Chiba 277-8561, Japan}

\author{M. Sprung}
\affiliation{Deutsches Elektronen-Synchrotron DESY, Notkestra$\beta$e 85, D-22607 Hamburg, Germany}

\author{N.L. Saini}
\affiliation{Dipartimento di Fisica, Universit{\'a} di Roma ``La Sapienza" - P. le Aldo Moro 2, 00185 Roma, Italy}

\date{\today}% It is always \today, today,

\begin{abstract}
We have used scanning micro x-ray diffraction to characterize different phases in superconducting K$_{x}$Fe$_{2-y}$Se$_2$ as a function of temperature, unveiling the thermal evolution across the superconducting transition temperature (T$_c\sim$32 K), phase separation temperature  (T$_{ps}\sim$520 K) and iron-vacancy order temperature (T$_{vo}\sim$580 K). In addition to the iron-vacancy ordered tetragonal magnetic phase and orthorhombic metallic minority filamentary phase, we have found a clear evidence of the interface phase with tetragonal symmetry. The metallic phase is surrounded by this interface phase below $\sim$300 K, and is embedded in the insulating texture. The spatial distribution of coexisting phases as a function of temperature provides a clear evidence of the formation of protected metallic percolative paths in the
majority texture with large magnetic moment, required for the electronic coherence for the superconductivity.  Furthermore, a clear reorganization of iron-vacancy order around the T$_{ps}$ and T$_c$ is found with the interface phase being mostly associated with a different iron-vacancy configuration, that may be important for protecting the percolative superconductivity in K$_{x}$Fe$_{2-y}$Se$_2$.
\end{abstract}

\maketitle
\section{Introduction}
The observation of superconductivity in the iron-based pnictides \cite{Hosono} and chalcogenides \cite{PNAS} has opened new frontiers in the field of layered materials with interesting interplay of atomic defects, magnetism, and superconductivity \cite{Johnston,Stewart}. In particular, defects in the iron-based chalcogenides are known to be important for the suppression of long-range magnetic order and appearance of the superconductivity \cite{PNAS,Johnston,TakRev}. Among these materials, the intercalated layered iron-chalcogenide system with chemical formula of A$_{x}$Fe$_{2-y}$Se$_2$  (A = K, Rb, Cs) ~\cite{Guo,Mizuguchi,Ying,Ming-Hu} is a good example in which a large magnetic moment is associated with the iron vacancy order ~\cite{Bao}. The A$_{x}$Fe$_{2-y}$Se$_2$ system also shows an intrinsic phase separation ~\cite{Ricci,Ricci1} and a delicate balance between an insulating magnetic phase associated with the iron vacancy order and a metallic phase considered to be superconducting below a transition temperature T$_c$ of $\sim$32 K. Indeed, A$_{x}$Fe$_{2-y}$Se$_2$ manifests peculiar microstructure, including iron vacancy order in the $ab$-plane, an antiferromagnetic order in the $c$-direction~\cite{Bao,Ryan} along with an intrinsic phase separation ~\cite{Ricci,Ricci1,Moss,Ryan,LiWei,Yuan,Wang2,Chen,SherMuSR} in which the majority phase with block antiferromagnetism has a stoichiometry of A$_{0.8}$Fe$_{1.6}$Se$_2$ (245) while the minority metallic phase is A$_{x}$Fe$_{2}$Se$_2$ (122). Incidentally, suppression of iron-vacancy order by high pressure produces a new phase with a T$_c$ of $\sim$56 K \cite{highP}. A variety of experimental techniques have been used to study the intrinsic phase separation \cite{Guo,Mizuguchi,Ying,Ming-Hu,Ryan,Bao,Ricci,Ricci1,LiWei,Yuan,Moss,Wang2,Chen,SherMuSR,Simonelli12,Oiwake13,Bendele14,Shoemaker12}, revealing a wealth of information on the peculiar microstructure of these materials.

While most of the studies on A$_{x}$Fe$_{2-y}$Se$_2$ have been focused either on the iron-vacancy ordering or the phase separation, there are limited efforts to address the relationship between the peculiar microstructure and the multi-band  electronic structure sustaining the superconductivity. Very recently, an orbital selective Mott phase (OSMP) has been proposed to have an important role in A$_{x}$Fe$_{2-y}$Se$_2$ \cite{osmp}. This phase has been observed around 100-300 K in the multi-band metallic phase by angle resolved photoemission spectroscopy (ARPES) \cite{Shen}. In addition, a recent high pressure study \cite{Zhao14} has underlined the importance of the OSMP phase as an intermediate phase between the iron-vacancy ordered insulating texture and the minority metallic phase. Also a recent high energy x-ray emission (XES) study has found anomalous evolution of the magnetic phase below 300 K that can be assigned to the OSMP phase \cite{Simonelli14}. These new findings suggest that a space resolved characterization of different phases is of upmost importance.

Space resolved diffraction is emerging as one of the key experimental tools to study the distribution of phases in intrinsically inhomogeneous materials, and has been efficiently exploited to obtain useful information on the structure-function relationship in a variety of systems \cite{FratNat,PocciaNmat,PocciaPRB,PocciaPNAS,CampiPRB,RicciSrep,RicciNJP}.
Earlier, we have used this space resolved micro x-ray diffraction ($\mu$XRD) to explore intrinsic nanoscale phase separation in K$_{x}$Fe$_{2-y}$Se$_{2}$~\cite{Ricci}. The temperature dependent study revealed the phase separation below $\sim$ 520 K and a $\sqrt{5}\times\sqrt{5}$ superstructure due to  iron-vacancy order below $\sim$ 580 K. The study also revealed spatial distribution of different phases, i.e.: i) the majority phase ($\sim$70-90\%) with $\sqrt{5}\times\sqrt{5}$  superstructure due to iron vacancy order and; ii) the minority phase ($\sim$10-30\%) with a compressed in-plane lattice.   In this work, we have further exploited the technique and focused on finding the possible interface phase in K$_{x}$Fe$_{2-y}$Se$_{2}$ by scanning $\mu$XRD in a wide temperature range, including the superconducting critical temperature (T$_c\sim$32 K), phase separation temperature  (T$_{ps}\sim$520 K) and iron-vacancy order temperature (T$_{vo}\sim$580 K). We have used a coherent x-ray source with microscopic spatial resolution to explore different coexisting phases and found that, in addition to the tetragonal majority antiferromagnetic insulating (AFM) phase and orthorhombic minority paramagnetic metallic filamentary (PAR) phase, there exists a phase having tetragonal symmetry, that appears below a temperature of $\sim$300 K and distributes at the interface of the two main phases. The space resolved diffraction across the superconducting transition temperature T$_c$ provides a clear evidence of percolative paths protected by the interface phase, required for the coherent electronic state for the superconductivity.  The results also reveal  reorganization of an iron-vacancy ordered phase around T$_c$, suggesting that the interface phase should be associated with a different iron-vacancy configuration and that may have an important role in the percolative superconductivity of K$_{x}$Fe$_{2-y}$Se$_2$. 

\section{Experimental details.}\label{}
The K$_x$Fe$_{2-y}$Se$_2$ single crystal samples were prepared using the Bridgman method~\cite{Mizuguchi}. After the growth, the single crystals were sealed into a quartz tube and annealed for 12 hours at 600$^\circ$C. The electric and magnetic characterizations were performed by resistivity measurements in a physical property measurement system (PPMS) (Quantum Design) and magnetization measurements in a superconducting quantum interference device (SQUID) magnetometer (Quantum Design). The samples exhibit a sharp superconducting transition at T$_{c}$ of $\sim$32 K. 

The scanning 
$\mu$XRD experiments were carried out at the Coherence Beamline P10 of PETRA III synchrotron Hamburg. The x-ray beam, produced by a 5m long undulator (U29)  is monochromatized by a cooled Si(111) double crystal monochromator. An x-ray energy of 8 KeV with a bandwidth of dE/E$\sim$1.4$\times$10$^{-4}$ was selected. 
This collimated coherent x-ray beam was focused using the beryllium refractive lens (CRL) transfocator to a size of about 2$\times$2 $\mu$m$^2$ on the sample positioned at 1.6 m down stream of the transfocator center. 
The incident flux on the sample was about 1-2$\times$10$^{11}$ photons/s. 
The exit window of the heating chamber and He-cryostat as well as the entrance window of the detector flight path was covered by a 25 $\mu$m thick Kapton sheet. The scattered signal was detected at a sample to detector distance of $\sim$5 m using a large horizontal scattering set-up. A PILATUS 300 K detector was used to record the 2$\times$2 $\mu$m$^2$ x-rays scattered by the sample. The intensity, I, of different phases was integrated over square subareas of the images recorded by the CCD camera in reciprocal-lattice units (r.l.u.) and then normalized to the intensity (I$_0$) of the tail of the main crystalline reflections at each point (x, y) of the sample reached by the translator. For the measurements, the sample was cooled to the lowest temperature and the measurements were performed in the heating cycle.

\section{Results and discussions}\label{}
%Figure 1
\begin{figure*}
\includegraphics[width=\linewidth]{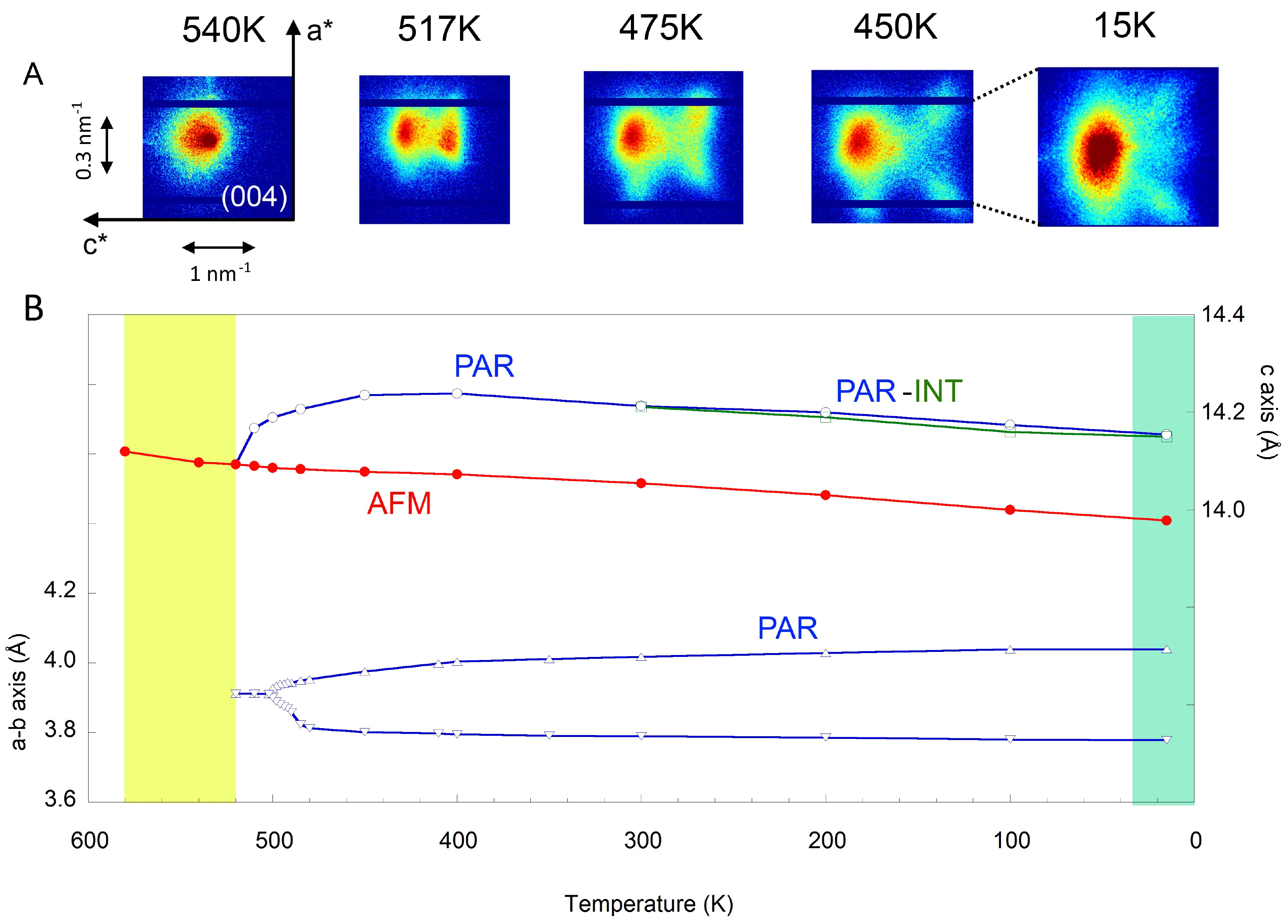}
\caption{(a) Temperature evolution of the (004) Bragg peak of superconducting K$_x$Fe$_{2-y}$Se$_2$ single crystal showing intrinsic phase separation at 520 K and a tetragonal to orthorhombic structural transition of the minority metallic phase below 500 K.
(b)	Temperature dependence of the lattice parameters for different phases. Color shades are used to show superconducting and tetragonal non-magnetic regions in (b).}
\label{1}
\end{figure*}

Here, we start with the already known intrinsic phase separation \cite{Ricci,Bao} in superconducting K$_x$Fe$_{2-y}$Se$_2$. Unlike the earlier studies, we have performed $\mu$XRD measurements in a wide temperature range using a coherent x-ray source.
Figure 1 (a) shows the temperature evolution of the (004) Bragg peak in the ac-plane, measured on a single crystal sample of superconducting K$_x$Fe$_{2-y}$Se$_2$. At 540 K, a sharp and highly symmetric peak appears due to a tetragonal structure (a=b=4.01 \AA, c=13.84 \AA, space group I4/mmm). Once the sample is cooled across the phase separation temperature of $\sim$520 K \cite{Ricci,Ricci1}, the peak splits in two, and a new tetragonal phase with elongated c-axis (see e.g. in Fig. 1, the (004) peak at 517 K) appears, coexisting with the main phase. Thus, K$_x$Fe$_{2-y}$Se$_2$ is phase separated containing a majority phase with $\sqrt{5}\times\sqrt{5}$ superstructure due to ordered iron vacancies and associated block antiferromagnetism \cite{Ricci,Bao} (AFM phase). The phase separation occurs due to thermal contraction affecting the iron vacancy ordering configuration with coupled magnetism in the main phase. Further cooling hardly affects the main phase, while the peak associated with the new phase (with elongated c-axis) reveals reduction in the crystallographic symmetry. Indeed, the peak splits diagonally indicating orthorhombic symmetry (see e.g. in Fig. 1, the profile at 450 K). This minority phase is known to be free from any iron-vacancy order \cite{Bao}, and is metallic (PAR phase). If the sample is cooled further, the AFM and PAR phases are hardly affected, however a new peak appears, reflecting the appearance of a third phase characterized by an elongated c-axis. We will come back to show that this third phase appears at the interface between the majority tetragonal AFM phase and minority orthorhombic PAR phase.  This new phase, having average tetragonal symmetry is called, the interface (INT) phase. 

The temperature evolution of the lattice parameters of different phases in K$_x$Fe$_{2-y}$Se$_2$ is shown in Fig. 1(b). The phase separation appears below $\sim$ 520 K in the AFM and PAR phases, however, the symmetry of the PAR phase changes by further cooling down to $\sim$ 500 K. In addition, the INT phase appears below $\sim$ 300 K, with c-axis similar to the PAR phase. Therefore, three phases coexist at low temperature. The average lattice parameters measured here are consistent with earlier studies on superconducting the K$_x$Fe$_{2-y}$Se$_2$ system \cite{Ricci,Ricci1,Bao,Mizuguchi,Shoemaker12}. 

%Figure 2
\begin{figure}
\includegraphics[width=\linewidth]{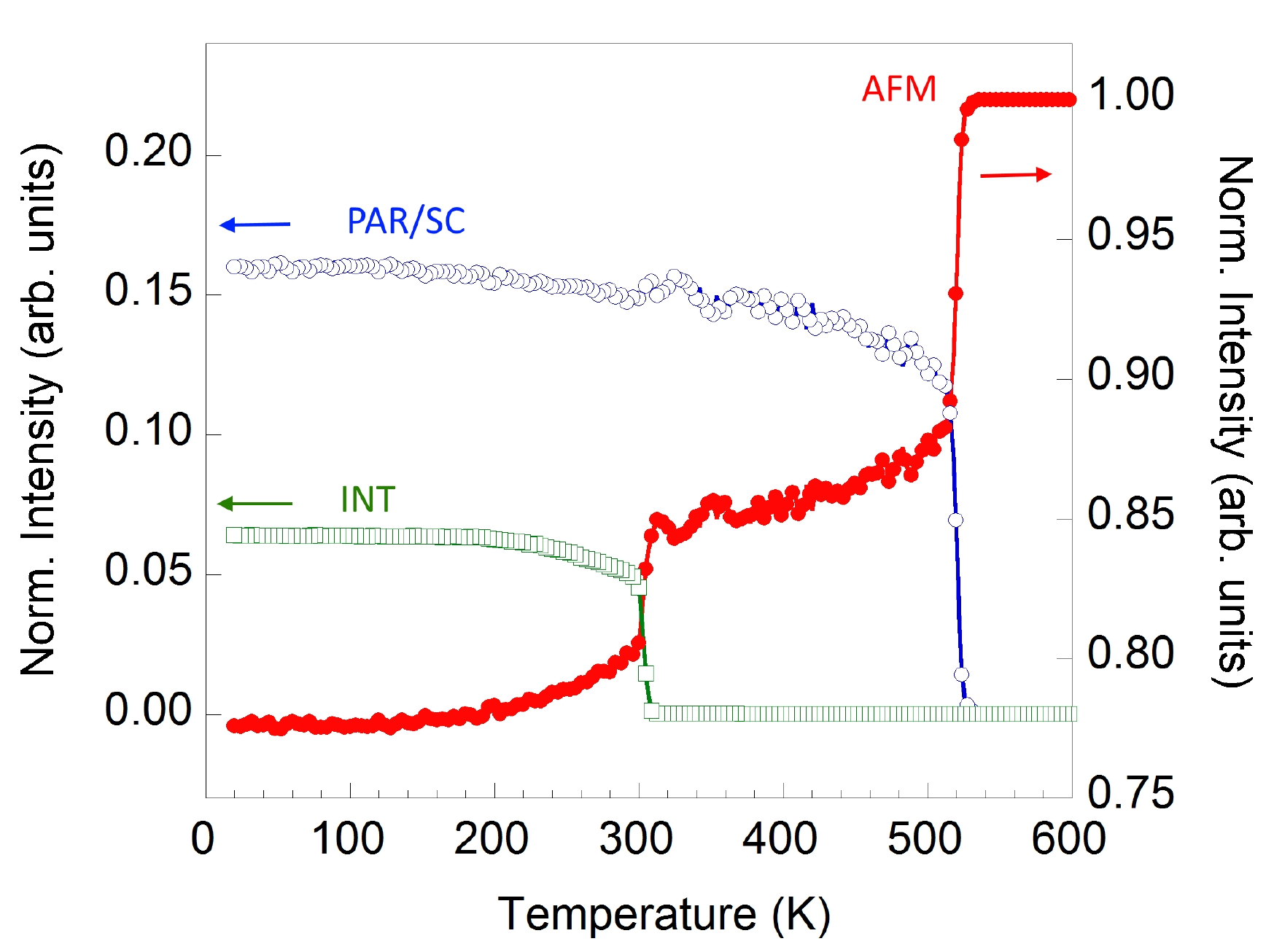}
\caption{
Normalized intensity of (004) peak of K$_x$Fe$_{2-y}$Se$_2$ for different phases as a function of temperature, representing the evolution of their relative weights. The majority phase decreases sharply (red) due to the appearance of the new phase (blue) below the phase separation temperature (T$_{ps}\sim$520 K). Further cooling results in the appearance of an interface phase (green) at $\sim$300 K.
}
\label{3}
\end{figure}

The relative weights of coexisting phases have been estimated by the intensity of the (004) Bragg peak corresponding to the different crystallographic phases. Figure 2 shows the normalized intensity of the (004) diffraction peak for different phases plotted as a function of temperature.
As expected, the 
majority AFM phase decreases sharply across the phase separation temperature ($\sim$520 K) due to the appearance of the minority PAR phase.
The majority AFM phase 
contributes $\sim$85-90\% while the remaining $\sim$10-15\% is the PAR phase 
at 300 K$\le$T$\le$520 K. This is consistent with earlier studies, estimating relative weights of the two phases at $\sim$80-90\% and $\sim$10-20\% respectively \cite{Bao,SherMuSR,Bendele14,Simonelli14}. While cooling across $\sim$300 K, 
the majority AFM phase suffers a further decrease  to about $\sim$78\% with the appearance of a third (INT) phase having maximum weight of $\sim$6\%. The PAR phase also appears to gain, 
reaching a value $\sim$16\%. Therefore, at low temperature (T$\le$300 K), there are three phases in K$_x$Fe$_{2-y}$Se$_2$ with different relative weights.

%Figure 3
\begin{figure}
\includegraphics[width=\linewidth]{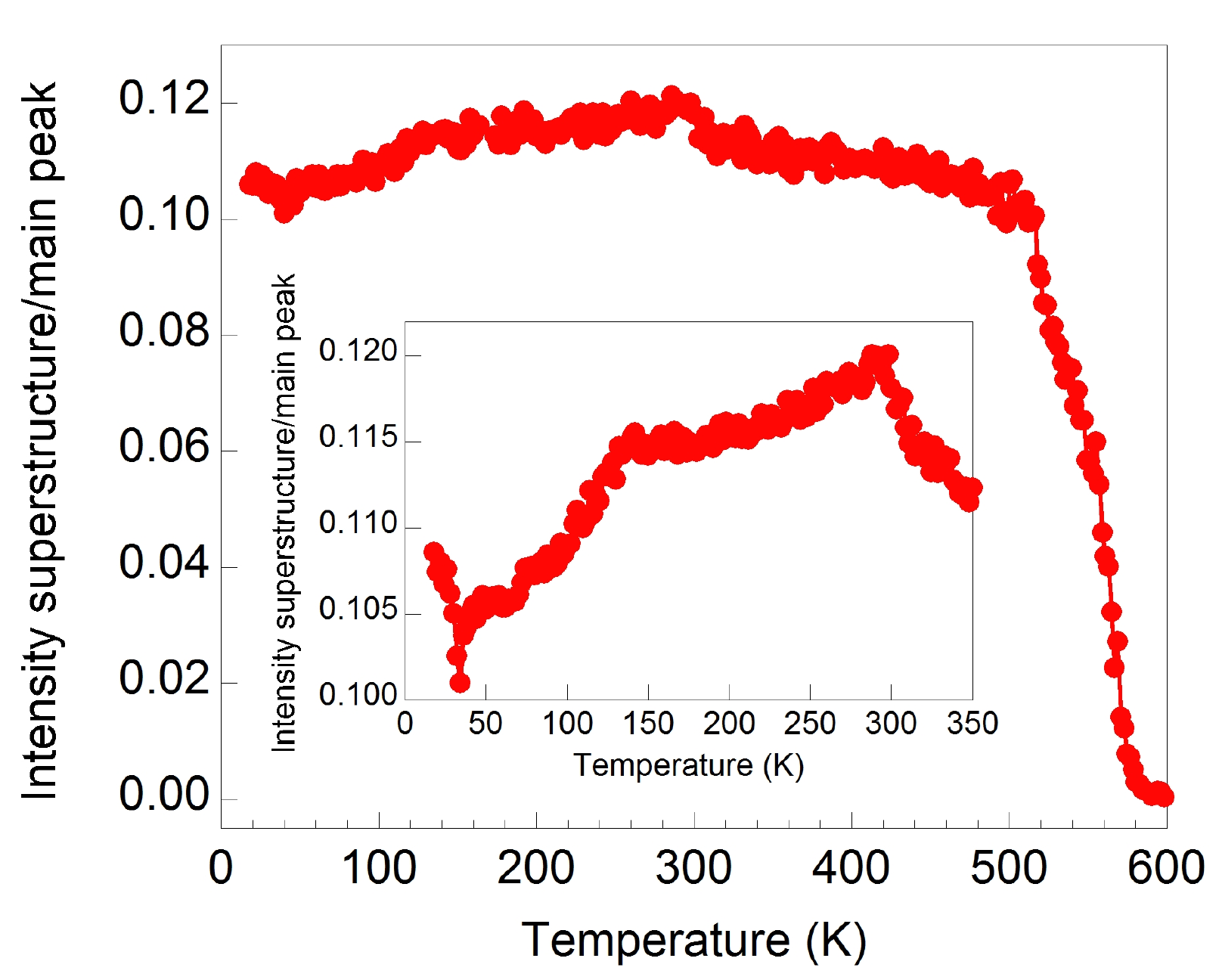}
\caption{
Intensity of the $\sqrt{5}\times\sqrt{5}$ superstructure peak 
normalized with respect to the corresponding (004) diffraction peak, is shown as a function of temperature. A sharp jump at $\sim$580 K due to iron-vancancy order 
is apparent. A small deviation at $\sim$520 K indicates changing microstructure due to phase separation. A zoom over the low temperature  is shown as an inset revealing anomalous change in the microstructure properties below 300 K, and at T$_c\sim$32 K.}
\label{4}
\end{figure}

Let us focus on the majority AFM phase that is characterized by the iron vacancy order and the associated antiferromagnetic order \cite{Bao}. To further explore the iron-vacancy order we have followed the temperature dependence of the  $\sqrt{5}\times\sqrt{5}$ superstructure peak normalized with respect to the corresponding (004) peak. This quantity is shown as a function of temperature in Fig. 3, revealing the degree of the vacancy order in the majority phase. The superstructure peak shows up below the iron-vacancy ordering temperature, apparent from the the sharp jump at $\sim$580 K. 
The normalized intensity shows a small deviation from the order parameter-like behavior at $\sim$520 K, expected due to change in the microstructure and associated iron-vacancy order configuration across the phase separation. Upon cooling further, 
a small upturn can be seen at $\sim$350 K  followed by a gradual decrease before a rapid decrease around $\sim$150 K due to evolving iron vacancy order configurations. This anomalous change is followed by a sharper (albeit small) decrease at T$_c\sim$32 K with an upturn at lower temperature. The anomalous evolution of the iron-vacancy ordered phase is apparent from the zoom over the low temperature range shown as the inset in Fig. 3. Such a thermal evolution suggests that iron-vacancy order is anomalously affected by cooling and should be related with the symmetry of different phases at local scale. Also, the anomaly around 
T$_c\sim$32 K indicates that the superconductivity  should be affected by the iron-vacancy order in this system.

%Figure 4
\begin{figure}
\includegraphics[width=8 cm]{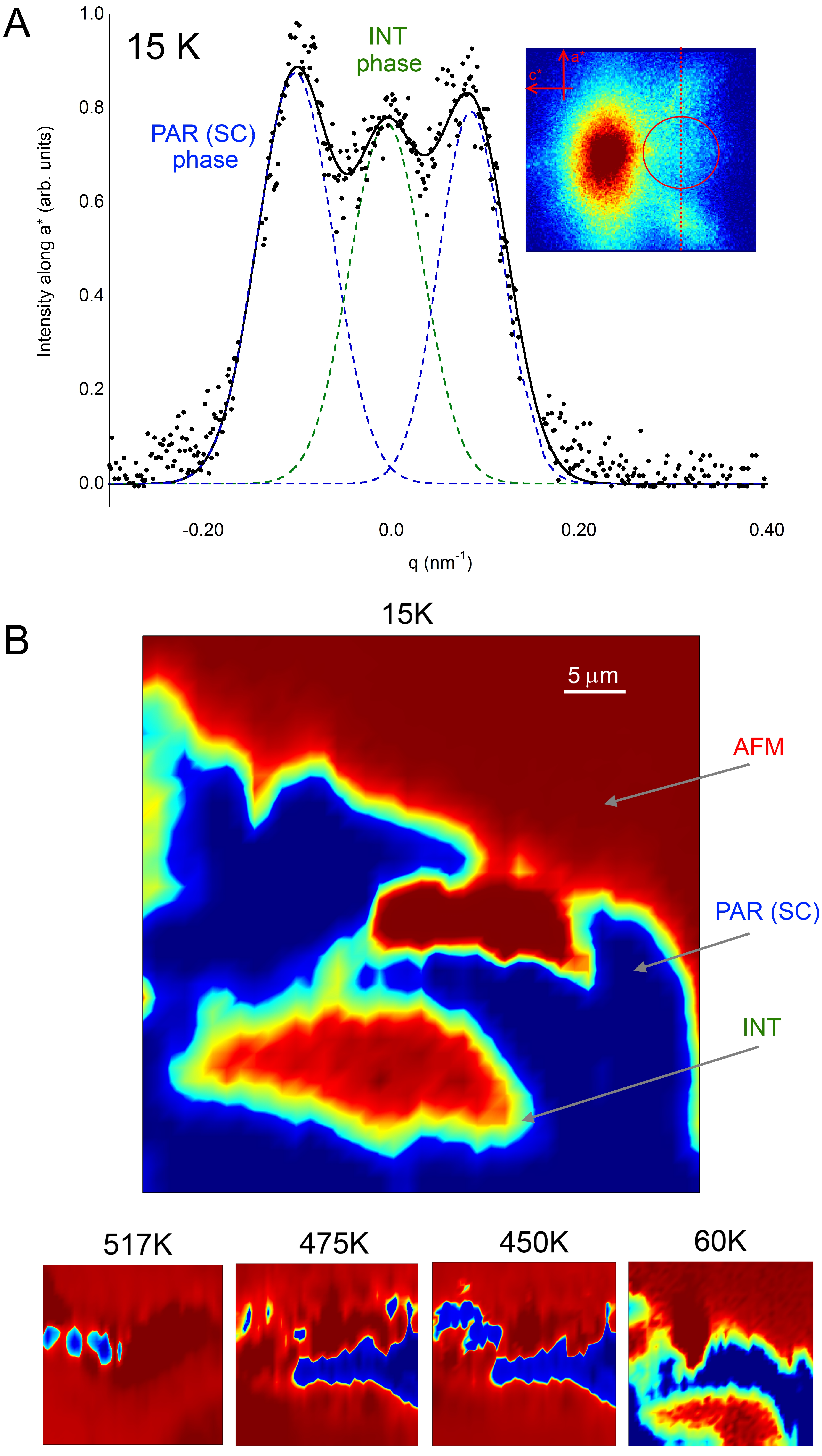}
\caption{
(a) The line profile of (004) diffraction peak (inset) along a* direction at 15 K for K$_x$Fe$_{2-y}$Se$_2$. The  line profile contains three peaks (3 Lorentzians and a model fit is also shown), with the central peak being due to the interface phase while the other two are due to the minority orthorhombic phase. 
(b)	Spatial distribution of different phases at 15 K for a selected area of 80$\times$80 $\mu$m. The interface phase (green) is clearly visible between the majority (red) and the minority (blue) phases. The lower panel show the spatial distribution of different phases at several temperatures, revealing evolution of percolative paths (blue) below the phase separation temperature, getting protection of the interface phase (green) at low temperature. 
}
\label{}
\end{figure}

Let us recall the known phase separation ~\cite{Ricci,Wang2,Chen,Simonelli12,Oiwake13,Bendele14,Shoemaker12}.  It is established that the superconducting K$_x$Fe$_{2-y}$Se$_2$ contains:  (i) a majority insulating 245-phase and; (ii) a minority metallic 122-phase~\cite{Ricci,Shoemaker12}. The 245 and the 122 phases have tetragonal and orthorhombic structures with the latter being slightly compressed in the plane and expanded in the out of plane ~\cite{Ricci,Ricci1}. Here, it is clear that, in addition to the majority 245 and minority 122 phase, an interface phase appears as a third phase at low temperature that may be important for the superconductivity. While such a  phase has been proposed on the basis of recent high pressure measurements
on A$_x$Fe$_{2-y}$Se$_2$~\cite{Zhao14}, we have observed it directly [see, e.g. Figure 4(a)]. This interface phase is characterized by a tetragonal symmetry with the c-axis longer than the one of the AFM phase (however, almost similar to the c-axis of the PAR phase).

To get further information on the INT phase and its spatial distribution, we have performed a limited area space resolved $\mu$XRD map as a function of temperature. The spatial distribution of different phases at 15 K is shown in Figure 4B. The intensity distribution is created by integrating intensities of (004) diffraction peaks corresponding to different phases
The majority and minority phases in the superconducting state (15 K) are visible as large disconnected regions and percolative paths. These paths are surrounded by a finite area INT phase. It is worth mentioning that the scan area has been selected randomly and the observation of the interface phase does not depend on the selected area for the $\mu$XRD image.

The temperature evolution of the overall spatial distribution of different phases is displayed in Figure 4B. The minority PAR phase appears below the phase separation temperature and increases with cooling, developing in percolative paths (blue). These paths are protected by the INT phase (green) below $\sim$300 K. Therefore, the INT phase indeed evolves below 300 K consistent with the recent XES study \cite{Simonelli14}. The INT phase is liklely to be the OSMP indicated in the ARPES study \cite{Shen}
where the Fe 3d $xy$ electrons are localized and magnetic. On the other hand, the phase separation in the majority insulating AFM phase and the minority metallic PAR phase is known to occur below $\sim$520 K, while the INT phase appears below $\sim$300 K. In this temperature range a resistivity hump also appears \cite{Mizuguchi,Shoemaker12}. Therefore, it is clear that the PAR phase (embedded in the majority AFM phase) is not enough for a metallic conductivity in the system, and hence the INT phase is required for the same and for the superconductivity on cooling across T$_c$. Furthermore, the results clearly show that the AFM phase is reduced by the metallic INT phase in establishing percolative paths. Thus, the reported spatial distribution of coexisting phases as a function of temperature is a clear evidence of the formation of protected metallic percolative paths in the majority AFM texture with large magnetic moment. This protection by the INT phase may help in realizing the superconducting coherent quantum state in the metallic filamentary phase for the percolative superconductivity in K$_x$Fe$_{2-y}$Se$_2$.

\section{Conclusions}\label{S:conclusions}
In summary, we have used space resolved micro x-ray diffraction measurements on superconducting K$_{x}$Fe$_{2-y}$Se$_2$ as a function of temperature. We find that, the phase separation at low temperature is characterized by the coexistance of the majority tetragonal magnetic phase, orthorhombic minority metallic phase, and an interface tetragonal phase appearing below $\sim$300 K. The results reveal an anomalous behavior of the iron-vacancy ordering, affected by phase separation and superconductivity. Spatial distribution of different coexisting phases measured by space resolved micro diffraction provides a clear evidence of the formation of percolative paths with decreasing temperature, having protection by the interface phase required for the electronic coherence for the superconductivity. The results suggest that the interface phase should be associated with different iron-vacancy configuration and likely to have an important role in the percolative superconductivity of K$_{x}$Fe$_{2-y}$Se$_2$.

\section*{Acknowledgments}
We thank the PETRA staff for the assistance during the measurements. Two of us (Y.M and T.M.) would like to acknowledge hospitality at the Sapienza University of Rome. One of us (N.P.) acknowledges financial support from the Marie Curie IEF project for career development. The work is partially supported by PRIN2012 (Grant No. 2012X3YFZ2) of MIUR, Italy.
%\end*{acknowledgments}


\begin{thebibliography}{40}

\bibitem{Hosono}Y. Kamihara, T. Watanabe, M. Hirano, H. Hosono, J. Am. Chem. Soc.  {\bf 130}, 3296 (2008).

\bibitem{PNAS} F.-C. Hsu, J.-Y. Luo, K.-W. Yeh, T.-K. Chen, T.-W.
Huang, P. M. Wu, Y.-C. Lee, Y.-L. Huang, Y.-Y. Chu, D.-C. Yan, M.-K.
Wu, Proc.  Nat.  Acad.  Sci.  {\bf 105}, 14262 (2008).

\bibitem{Johnston} D.C. Johnston Advances in Physics {\bf 59} 803 (2010).

\bibitem{Stewart} G.R. Stewart Rev. Mod. Phys. {\bf 83} 1589 (2011)

\bibitem{TakRev} Y. Mizuguchi and Y. Takano J. Phys. Soc. Japan {\bf 79} 102001 (2010).

\bibitem{Guo}%
J. Guo, S. Jin, G.Wang, S.Wang, K. Zhu, T. Zhou, M. He, and X. Chen, 
Phys. Rev. B {\bf 82}, 180520 (2010). 

\bibitem{Mizuguchi}%
Y. Mizuguchi, H. Takeya, Y. Kawasaki, T. Ozaki, S. Tsuda, T. Yamaguchi, and Y. Takano,,  Appl. Phys. Lett. {\bf 98}, 042511 (2011).

\bibitem{Ying}%
J. J. Ying, X. F. Wang, X. G. Luo, A. F. Wang, M. Zhang, Y. J. Yan, Z. J. Xiang, R. H. Liu, P. Cheng, G. J. Ye, and X. H. Chen, Phys. Rev. B {\bf 83}, 212502 (2011). 

\bibitem{Ming-Hu}%
M.H. Fang, H.D. Wang, C.H. Dong, Z.J. Li, C.M. Feng, J. Chen, and H. Q. Yuan, Europhys. Lett. {\bf 94}, 27009 (2011). 

\bibitem{Bao}%
W. Bao, Q.Z. Huang,  G.F. Chen, M. A. Green,  D.M. Wang,  J.B. He, and  Y.-M. Qiu, Chin. Phys. Lett. {\bf 28}, 086104 (2011).

\bibitem{Ricci}%
A. Ricci, N. Poccia, G. Campi, B. Joseph, G. Arrighetti, L. Barba, M. Reynolds, M. Burghammer, H. Takeya, Y. Mizuguchi, Y. Takano, M. Colapietro, N. L. Saini, and A. Bianconi, Phys. Rev. B {\bf 84}, 060511(R) (2011); 

\bibitem{Ricci1}%
A. Ricci, N. Poccia, B. Joseph, G. Arrighetti, L. Barba, J. Plaisier, G. Campi, Y. Mizuguchi, H. Takeya, Y. Takano, N.L. Saini and A. Bianconi, Superconductor Science and Technology 24, 082002 (2011).

\bibitem{Moss} V. Ksenofontov, G. Wortmann, S. A. Medvedev, V. Tsurkan, J. Deisenhofer, A. Loidl, and C. Felser, Phys. Rev. B {\bf 84}, 180508 (2011).

\bibitem{Ryan}%
D.H. Ryan, W. N. Rowan-Weetaluktuk, J.M. Cadogan, R. Hu, W.E. Straszheim, S.L. Bud'ko, and P.C. Canfield, Phys. Rev. B {\bf 83}, 104526 (2011).

\bibitem{LiWei} Wei Li, Hao Ding, Peng Deng, Kai Chang, Canli Song, Ke He, Lili Wang, Xucun Ma, Jiang-Ping Hu, Xi Chen, Qi-Kun Xue, Nature Physics {\bf 8}, 126 (2012).

\bibitem{Yuan} R.H. Yuan, T. Dong, Y.J. Song, P. Zheng, G.F. Chen, J.P. Hu J.Q. Li, N.L. Wang, Scientific Reports 2 221 (2012).

\bibitem{Wang2}%
Z. Wang, Y. J. Song, H. L. Shi, Z. W. Wang, Z. Chen, H. F. Tian, G. F. Chen, J. G. Guo, H. X. Yang, and J. Q. Li, Phys. Rev B {\bf 83}, 140505(R) (2011).

\bibitem{Chen}%
F. Chen, M. Xu, Q. Q. Ge, Y. Zhang, Z. R. Ye, L. X. Yang, Juan Jiang, B. P. Xie, R. C. Che, M. Zhang, A.
F. Wang, X. H. Chen, D. W. Shen, J. P. Hu, and D. L. Feng, Phys. Rev. X {\bf 1}, 021020 (2011).

\bibitem{SherMuSR}Z. Shermadini,  H. Luetkens, R. Khasanov, A. Krzton-Maziopa, K. Conder,  E. Pomjakushina, H-H. Klauss, and A. Amato, Phys. Rev. B {\bf 85}, 100501 (2012). 

\bibitem{highP} L. Sun, X.-J. Chen, J. Guo, P. Gao, Q.-Z. Huang, H. Wang, M. Fang, X. Chen, G. Chen, Q. Wu, C. Zhang, D. Gu, X. Dong, L. Wang, K. Yang, A. Li, X. Dai, H.-k. Mao, and Z. Zhao, Nature (London) {\bf 483}, 67 (2012).

\bibitem{Simonelli12}%
L. Simonelli, N. L. Saini, M. Moretti Sala, Y. Mizuguchi, Y. Takano, H. Takeya, T. Mizokawa, and G. Monaco, Phys. Rev. B {\bf 85}, 224510 (2012).

\bibitem{Oiwake13}%
M. Oiwake, D. Ootsuki, T. Noji, T. Hatakeda, Y. Koike, M. Horio, A. Fujimori, N. L. Saini, and T. Mizokawa, Phys. Rev. B {\bf 88}, 224517 (2013).

\bibitem{Bendele14}%
M. Bendele, A. Barinov, B. Joseph, D. Innocenti, A. Iadecola, H. Takeya, Y. Mizuguchi, T. Takano, T. Noji, T. Hatakeda, Y. Koike, M. Horio, A. Fujimori, D. Ootsuki, T. Mizokawa, and N. L. Saini, Sci. Rep. {\bf 4}, 5592 (2014).

\bibitem{Shoemaker12}%
D. P. Shoemaker, D. Y. Chung, H. Claus, M. C. Francisco, S. Avci, A. Llobet, 
and M. G. Kanatzidis, Phys. Rev. B {\bf 86}, 184511 (2012).

\bibitem{osmp} Rong Yu and Qimiao Si, Phys. Rev. Lett.  {\bf 110} 146402 (2013).

\bibitem{Shen}%
M. Yi, D. H. Lu, R. Yu, S. C. Riggs, J.-H. Chu, B. Lv, Z. K. Liu, M. Lu, Y.-T. Cui, M. Hashimoto, S.-K. Mo, Z. Hussain, C.W. Chu, I. R. Fisher, Q. Si, and Z.-X. Shen, Phys. Rev. Lett. {\bf 110}, 067003 (2013).

\bibitem{Zhao14}%33
P. Gao, R. Yu, L. Sun, H. Wang, Z. Wang, Q. Wu, M. Fang, G. Chen, J. Guo, C. Zhang, D. Gu, H. Tian, J. Li, J. Liu, Y. Li, X. Li, S. Jiang, K. Yang, A. Li, Q. Si, and Z.X. Zhao, Phys. Rev. B {\bf 89}, 094514 (2014).

\bibitem{Simonelli14} L. Simonelli, L. Simonelli, T. Mizokawa, M. Moretti Sala, H. Takeya, Y. Mizuguchi, Y. Takano, G. Garbarino, G. Monaco, and N.L. Saini, Phys. Rev. B {\bf 90}, 214516 (2014).

\bibitem{FratNat} M. Fratini, N. Poccia, A. Ricci, G. Campi, M. Burghammer, G. Aeppli, and A. Bianconi, Nature 466, 841 (2010).

\bibitem{PocciaNmat} N. Poccia, M. Fratini, A. Ricci, G.  Campi, L. Barba, A. Vittorini-Orgeas, G.  Bianconi, G. Aeppli and A. Bianconi, Nature Materials 10, 733 (2011)

\bibitem{PocciaPRB} N. Poccia, G. Campi, M. Fratini, A. Ricci, N. L. Saini, and A. Bianconi, Physical Review B 84, 100504 (2011).

\bibitem{PocciaPNAS} N. Poccia, A. Ricci, G. Campi M. Fratini, A. Puri, D. Di Gioacchino A. Marcelli, M. Reynolds, M. Burghammer N.L. Saini,  G. Aeppli and A. Bianconi Proceedings of the National Academy of Sciences, 109, 15685 (2012). 

\bibitem{CampiPRB}  G. Campi, A. Ricci, N. Poccia, L. Barba, G. Arrighetti, M. Burghammer, A. S. Caporale, and A. Bianconi, Physical Review B 87, 014517 (2013).

\bibitem{RicciSrep}  A. Ricci, N. Poccia, G. Campi, F. Coneri, A. S. Caporale, D. Innocenti, M. Burghammer, M. Zimmermann, and A. Bianconi, Scientific Reports 3, 2383 (2013).

\bibitem{RicciNJP}  A. Ricci, N. Poccia, G. Campi, F. Coneri, L. Barba, G. Arrighetti, M. Polentarutti, M. Burghammer, M. Sprung, Martin, et al., New Journal of Physics 16, 053030 (2014).

\end{thebibliography}
\end{document}